\documentclass[12pt,a4paper]{article}
\usepackage{amsmath}
\usepackage{cite}
\usepackage{graphicx}
\usepackage{amsfonts}
\usepackage{amssymb}
\def\my6j#1#2#3#4#5#6{
\left\{
\begin{array}
[c]{cc}%
#1 & #3\\
#2 & #4%
\end{array}
\mid
\begin{array}{c}
#5\\
#6%
\end{array}\right\}_{b}}

\def\fussu2#1#2#3#4#5#6{
F^{SU(2)}_{#5#6}\left[
\begin{array}
[c]{cc}%
#3 & #2\\
#4 & #1%
\end{array}
\right]}

\def\sl2{
\hat{sl}(2)}

\begin{document}
\bigskip
{\flushright
\hfill\hbox{SPhT-T02/135}}  {\flushright
\hfill\hbox{AEI 2002-069}}
\vspace{2cm}

\begin{center}
{\Large \textbf{Are there really any $AdS_2$ branes}} \\
\vspace{0.5cm} {\Large\textbf{in the euclidean (or not) $AdS_3$?}}\\

\vspace{1.2cm} {\Large B\'en\'edicte Ponsot,\footnote{\textsf
{bponsot@aei-potsdam.mpg.de, \ ponsot@spht.saclay.cea.fr}}\ \ \ Sebasti\'an
Silva\footnote{\textsf {silva@aei-potsdam.mpg.de}}} \vspace{0.8cm}

{\it $^{1,2}$Max Planck Institut f\"ur Gravitationsphysik, Albert
Einstein Institut, \\
Am M\"uhlenberg 1, D-14476 Golm, Germany.} \vspace{0.5cm}

 {\it $^{1}$Service de Physique Th\'eorique, Commissariat \`a l'\'energie
atomique,\\
 CEA L'orme des merisiers,
 F-91191 Gif sur Yvette, France.}
\end{center}
\begin{abstract}
We do not find any $AdS_2$ branes, neither in the $H_3^+$ WZNW model
nor in the $SL(2,\mathbb{R})$ WZNW model. We then reexamine the case of
the branes that possess a $su(2)$ symmetry: we speculate that they would have to
live on the boundary of $AdS_3$. This cannot be realized in an
euclidean spacetime, but in the $SL(2,\mathbb{R})$ WZNW model by
analytical continuation.

\end{abstract}

\vspace{1cm}
\section*{Introduction}

The discussion of maximally symmetric branes in the euclidean (and
lorentzian) $AdS_3$ has received some attention recently
\cite{Sta,BaPe,PeRi,Ri,Ba,RaRo,HiSu,PS,LOPT,GKS,LOP,PST}. It seems
established that the possible maximally symmetric branes of the
model possess either a $sl(2,\mathbb{R})$ symmetry ($AdS_2$
branes) or a $su(2)$ symmetry (called spherical branes in
\cite{PST}), according to the different possible gluing conditions
for the currents on the boundary of the worldsheet. Using
conformal field theory techniques already developped for Liouville
field theory in \cite{FZZ,ZAZA}, the authors of \cite{GKS,LOP,PST}
proposed a microcospic description of these branes. In \cite{GKS}
were proposed one point functions in the $sl(2,\mathbb{R})$ and
$su(2)$ cases, which, as noticed in \cite{PST}, turned out to be
incorrect as the authors had wrong ansatz for its space time
dependence. In \cite{LOP}, the one point function in the
$sl(2,\mathbb{R})$ case was proposed, in \cite{PST}, the one
 point functions in the $sl(2,\mathbb{R})$ (which coincides with the one of \cite{LOP}),
  and  $su(2)$ cases
  as well as the boundary reflection amplitude were constructed. The Cardy condition
  was checked in both cases,
   leading to a continuous spectrum of boundary fields in the $sl(2,\mathbb{R})$ case,
    and to a discrete and finite
    one in the $su(2)$ case.\\
However, these results do not completly lie on solid
grounds, as the functional relations satisfied by
 these one point functions and the boundary reflection amplitude
  were only partially solved (see the conclusion
  of \cite{PST}). We would like to emphasize that {\it all} functional
relations {\it must} be solved. Thus, we do not
  consider the respective Cardy conditions as proven, as neither the quantities
   needed in the closed
   string channel (the one point function) nor in the open string channel
    (the boundary reflection amplitude in the
  $sl(2,\mathbb{R})$ case)
    are perfectly under control.
Some time ago, one of the authors of the present paper
 (B.P.) has emitted some doubts about the existence of these $AdS_2$ branes,
  (see the conclusion of
 \cite{P}), as it does not seem to be possible to construct
coherently the boundary three point function ({\it i.e.} the
scattering amplitude of open string states). So we decide to
reexamine the problem once more.
 This paper is organized as follows: section one contains basic definitions and notations, as well as the
  relations taken from \cite{PST} that define the $AdS_2$ and the $su(2)$-branes.
   In section two we check in the $AdS_2$ branes case the variational principle: it is satisfied
   without the need of adding any boundary action, so we do not find any source that renders the
    boundary problem interactive. This is problematic, for if the branes are curved, then it should be thanks to some boundary potential. It remains only the case of the straight brane to study (which is not curved by definition, and for which there is no boundary potential), but even in this case we do not manage to check the results of \cite{PST} against the second factorization constraint; so we propose to discard the existence of the
   $AdS_2$ branes. In section three,
    we reexamine the validity of the factorization constraints in the $su(2)$ case. They have no solution
    in an euclidean spacetime, but it might be possible to construct a consistent boundary
    conformal field theory in the $SL(2,\mathbb{R})$ WZNW model (the results are those of \cite{PST}, once the
     analytical continuation from euclidean to lorentzian of the spacetime
      and worldsheet coordinates is performed).
    A striking similarity with the boundary conditions that appear
     in Liouville field theory
     on the euclidean $AdS_2$ considered by Zamolodchikov
    and Zamolodchikov in \cite{ZAZA} is discussed.

\section{Preliminaries \cite{PST}}
The symmetric space $H_3^+$ consists of
hermitian $2 \times 2$ matrices $h$ with determinant $\det h =
1$ and positive trace. We parametrize this
space through coordinates $(\phi, \gamma,\bar{\gamma})$ such that
\begin{equation}\label{pggpar}
 h \ = \ \left( \begin{array}{cc} e^\phi \ & \ e^\phi \bar{\gamma} \\
                   e^\phi \gamma \ & \  e^\phi \gamma \bar{\gamma} + e^{-\phi}
                  \end{array} \right) \ \ .
\end{equation}
$\phi$ is real and $\gamma$ is
a complex coordinate with conjugate $\bar{\gamma}$.
The space $H_3^+$ is equipped with the following metric and
$H$-field,
\begin{eqnarray}
ds^2 =  \, d\phi^2 + e^{2 \phi} \, d\gamma d \bar{\gamma} \ \ \ ,
\ \ \ H  =  2 \, e^{2 \phi} \, d\phi \wedge d \bar{\gamma} \wedge
d \gamma \ \ .
\end{eqnarray}
We shall consider the following 2-form potential $B'$ for $H$: $$
B' \ = - e^{2 \phi} \, d \gamma \wedge d\bar{\gamma} \ \ . $$ The
B-field is imaginary so the theory is non-unitary. The action
functional for closed strings moving on $H_3^+$ then
reads\footnote{We correct here a misprint of \cite{PST}.}:
\begin{equation} \label{Saction}
 S(\phi,\gamma,\bar{\gamma}) \ = \ \frac{k}{\pi} \, \int \ dz \,
    d\bar z \ \left(
    \partial \phi \bar{\partial} \phi + e^{2 \phi} \, \bar{\partial} \gamma \partial \bar{\gamma}
     \right) \ \ .
\end{equation}
\paragraph{\it The currents.}
Let us introduce the following matrices
\begin{equation}
T_+ \ =\ \left(
\begin{matrix} 0 & -1\\ 0 & 0 \end{matrix}
\right),\quad
T_- \ =\ \left(
\begin{matrix} 0 & 0\\ 1 & 0 \end{matrix}
\right),\quad
T_0 \  = \ \frac{1}{2}\left(
\begin{matrix} 1 & 0\\ 0 & -1 \end{matrix}
\right).
\end{equation}
These are matrix representatives of the Lie algebra $sl(2,\mathbb{R})$,
{\it i.e.} \ they obey the relations $[T_0,T_\pm] = \pm T_\pm$ and
$[T_-,T_+] = 2 T_0$. For the chiral currents we use
$$ J(\bar z) \ := \ k \, h^{-1} \bar{\partial} h \ \ \quad \ \ \
   \bar{J}(z) \ := \ - k \, \partial h\, h^{-1} \ . $$
When we expand them according to $J (\bar z) = T_+ J^+ + T_- J^- +
2 T_0 J^0$, we obtain expressions for the components
\begin{eqnarray}
J^-(\bar z) & := &  k\, e^{2 \phi} \, \bar{\partial} \gamma \\[2mm]
J^0(\bar z) & := &  k\, \left(\bar{\partial}\phi -  e^{2\phi}\,
\bar{\gamma} \,
              \bar{\partial} \gamma \right) \\[2mm]
J^+(\bar z) & := & k\, \left( \bar{\gamma}^2 \, e^{2 \phi}\,
\bar{\partial} \gamma -
           \bar{\partial} \bar{\gamma} - 2 \, \bar{\gamma}\, \bar{\partial} \phi \right) \ \ .
\end{eqnarray}
The components of the anti-holomorphic currents are
constructed in an analogous way. Both sets of currents are
related by complex conjugation $ (J^\pm)^* = (\bar J)^{\mp}$
and $ (J^0)^* = - \bar J^0$.

\paragraph{\it The $AdS_2$ branes.} They correspond to
surfaces which are characterized by the equation
$$  tr \left( \begin{array}{cc} 0 & 1 \\ 1 & 0 \end{array}
     \right)h   = \ c \ .
 $$
where $c$ is a constant.
In terms of the coordinates introduced above one gets the
equation
\begin{eqnarray}
  e^\phi \, ( \gamma  + \bar{\gamma} ) \ = \ c \ .
\label{ads}
\end{eqnarray}
The currents satisfy the following relations on the boundary of
the worldsheet\footnote{Same as above.}
\begin{equation}\label{glue1}
 J^{\pm}(\bar{z}) \ = \ -\bar J^{\mp}(z)  \ \ \ , \ \ \
   J^0(\bar{z}) \ = \ -\bar {J^0} (z) \ \ .
\end{equation}
This implies that the current obeys $(J^\pm)^* = -J^\pm$ and
$(J^0)^* = J^0$ at $z = \bar z$.
\medskip
\paragraph{\it Branes that preserve a $su(2)$ symmetry.} They are
such that
$$   tr \left( \begin{array}{cc} 1 & 0 \\ 0 & 1 \end{array}
     \right) \ h \ = \ c \ \ .$$
with $c$ constant. This equation can be rewritten as:
\begin{eqnarray}
 e^{\phi} (\gamma \bar \gamma + 1) + e^{- \phi} \ = \ c.
\label{cond}
\end{eqnarray}
The currents satisfy\footnote{Same as above.} along the boundary
$z = \bar{z}:   J^{\pm} \ = \ \bar J^{\pm}  ,   \ J^0 \ = \ \bar
J^0. $ So we have $(J^\pm)^* = J^\mp , \ (J^0)^* = -J^0$, {\it
i.e.} a $su(2)$ current algebra on the boundary of the worldsheet.

\section{$AdS_2$ branes}
The star conditions for the currents give the following boundary
conditions for the fields at $z=\bar{z}$:
\begin{eqnarray}
(\partial-\bar {\partial}) \phi &=& - c e^{\phi } \bar {\partial} \gamma \ , \\
\gamma + \bar {\gamma } &=& c e^{-\phi } \ ,\\
\partial \bar {\gamma }+ \bar {\partial} \gamma  &=& 0 \ .
\end{eqnarray}
If one sets:$\ \ z = \tau +i\sigma, \ \ \bar{z} = \tau -i\sigma \
, \ \
\partial_{z} = \frac{1}{2} (\partial_{\tau} - i\partial_{\sigma}) \ ,
\ \
\partial_{\bar {z}} = \frac{1}{2} (\partial_{\tau} +
i\partial_{\sigma})\ $, then one can rewrite the boundary
conditions as
\begin{eqnarray}\label{bdcop}
i\partial_{\sigma } \phi &=& \frac{c}{2} e^{\phi }
\left(\partial_{\tau
}+i\partial_{\sigma } \right) \gamma \ ,  \\
\gamma + \bar {\gamma } &=& c e^{-\phi} \ ,\\
i \partial_{\sigma } \left( \gamma - \bar {\gamma }\right) &=& -c
\partial_{\tau } e^{-\phi }\ .
\end{eqnarray}
Then, using these conditions, the variational principle states
that at $\sigma =0$,
\begin{equation}\label{varprin}
\delta \phi \partial_{\sigma } \phi +\frac{1}{2} e^{2\phi }
\left(\delta \gamma (\partial_{\sigma } +i \partial_{\tau })
\bar{\gamma } + ( \partial_{\sigma } - i \partial_{\tau } ) \gamma
\delta \bar{\gamma }\right) = 0.
\end{equation}
The variational principle is thus satisfied {\it without the need
of adding any boundary term in the action.} This absence of boundary potential leads to some problems with respect to the analysis of \cite{PST}: it suggests that the observables depend on the bulk cosmological constant only (called $\lambda_b$ in \cite{PST}), whereas the one point function and boundary two point function proposed in \cite{PST} behaves like $\lambda_b^{\alpha}f(\cos\pi b^2(2\rho+1))$ where $\rho$ is the boundary condition and $\alpha$ some exponent. This scaling is actually what we would have expected had we found a boundary potential of the form $\sqrt{\lambda_b}\cos\pi b^2(2\rho+1)\int_{\mathbb{R}} dx B(x)$ (the real axis is the boundary of the worldsheet). Of course this argument is not sufficient to exclude the particular case of the straight brane for which $c=0$ ($\cos\pi b^2(2\rho+1)\equiv 0$ in this case): the scaling of \cite{PST} matches the expected scaling; however, the computation in this particular case shows that\footnote{One could object that the singularity of the conformal blocks at $z=x$ should be treated properly, which may be not straightforward. However, in the case of the $su(2)$-branes of the next section, such similar equations are solved so nicely that we are not convinced the problem lies in mathematics only.} the one point function (amongst others) proposed in
\cite{LOP,PST}  does not satisfy the factorization constraint arising when one
considers the degenerate field with spin $1/2b^2$.
We do not see how to construct a coherent conformal field theory in this case.

\section{$su(2)$-branes}
Let us remind that to construct the one point function in this
case, one first starts with a bulk two point function, where one
the fields has a spin $1/2$, the other a spin $j$ (see \cite{ZAZA}
for such a discussion). There are two equivalent ways of
expressing this two point function: either one first merges the
two operators in the bulk (this is given by a special case of the
three point function structure constant), then one approaches the
resulting field to the boundary, which is given by the one point
function; in the other channel, the two fields in the bulk do
approach the boundary, which gives a product of two one point
functions $U_{\rho}(1/2)U_{\rho}(j)$ ($\rho$ is the boundary
label). It is not clear whether the validity condition for such a
factorization has been properly discussed in \cite{GKS,PST}: one
requires a two point function to factorize into a product of two
one-point
 function, but, as mentionned in \cite{ZAZA}, this can be the case only when the fields are very far apart
%If we try to solve the relations for the currents in terms of the fields, then by making use of the relation
%$$ e^{\phi} (\gamma \bar \gamma + 1) + e^{- \phi} \ = \ c.
%$$
%we find that we should have $(\partial-\bar{\partial})\gamma=(\partial-\bar{\p%artial})\bar{\gamma}=0$,
%as well as $\bar{\gamma}\bar{\partial}\gamma+\gamma\partial\bar{\gamma}=0$. This implies that the product
 %$\gamma\bar{\gamma}$ should satisfy Dirichlet conditions on the boundary, as well as the field $\phi$.
 in the {\it spacetime} when they approach the boundary;
 to have the geodesical distance between
  the fields become infinite as they approach the boundary of the worldsheet, the fields should be at the
   boundary of $H_3^+$, {\it i.e.} at $\phi = +\infty$, and consequently $\gamma \bar \gamma + 1=0$ for
    the relation written in (\ref{cond}) to be satisfied \footnote{We take here the opportunity to mention
    that once again, the boundary conditions for the fields make the variational principle satisfied
     {\it without the need of extra
      boundary term in the action.}}. This equation does not have any solution in the
     euclidean spacetime, so it seems that these $su(2)$-branes do not exist
      in $H_3^{+}$\footnote{We understand that this discussion is somewhat speculative and should be
      handled with care: the requirement
     that the two point function decays into a product
of one point functions
     when the fields are taken very far apart is very formal here, as it holds in principle in {\it unitary} quantum field
      theories. In the case of Liouville field
      theory on the pseudosphere \cite{ZAZA}, it can be used {\it a priori} as this theory is believed to
      be a unitary conformal field theory, but it turns out {\it a posteriori} that the correlation functions
      in the bulk grow exponentially with the geodesic distance, which is certainly not an expected feature for a unitary conformal field theory. One could object
      the use of such an argument in the $H_3^+$ model, which is known to be a {\it non unitary} model.
       However, it might be that
      it is the non unitarity of the model that prevents the construction of  any D-branes.}. However,
      there would be a solution
in the $SL(2,\mathbb{R})$ model, where $\gamma, \bar{\gamma}$ are
substituted by real fields $a,b$ \footnote{The worldsheet also
becomes lorentzian.}. In this case, the boundary conformal field
theory constructed in \cite{PST} seems to be coherent (it is
straightforward to check that the one point function given in the
equation (3.41) of this paper indeed satifies {\it all} the
factorization constraints~-~only one factorization constraint
involving the degenerate field with spin $j=1/2$ was solved in
\cite{PST}). We would like to point out that the relation
$$
e^{\phi}\sim \frac{c}{1+ab},
$$
valid at $1+ab=0\ , \ \phi=+\infty$, is very reminiscent of what
was found for Liouville field theory on the pseudosphere
(euclidean $AdS_2$) \cite{ZAZA}. In this case the metric on the
Lobachevski plane is
$$
ds^2=e^{\phi_{L}(z,\bar{z})}|dz|^2
$$
where $\phi_L$ is the Liouville field, and
$$
e^{\phi_{L}(z,\bar{z})}=\frac{4R^2}{(1-z\bar{z})^2},
$$
$R$ is interpreted here as the radius of the pseudosphere. It was
shown in \cite{ZAZA} that the boundary fields live on the boundary
of the surface parametrized by $z\bar z =1$ (called the absolute),
and that the possible boundary conditions are in one
 to one correspondence with the degenerate representations of the Virasoro algebra.
We believe it is no accident if the boundary three point function of $SL(2,\mathbb{R})$ model can be written
 in terms of the boundary three point function in Liouville field
  theory \footnote{As explained in \cite{P}, the boundary three point function is constructed following the lines of
  \cite{PT3}:
   the normalization of the boundary operators can be found in \cite{PST}, equ. 4.40, and the fusion matrix
    is given by equ. 26 of \cite{P} (the parameter $-b^{-2}$ in this formula should be substituted by $b^{-2}$,
     and the superscript $MM$ that stands for minimal model should be replaced by LFT for Liouville field theory.}
(we discard on both sides the worldsheet and space-time dependence, and consider only the
structure constants).
$$
C_{j_3,j_2,j_1}^{\rho_3,\rho_2,\rho_1}=\frac{\Gamma(2+b^{-2}+j_{3}+j_2+j_1)}{
\Gamma(2+b^{-2}+j_{3}+j_2+j_1-m)}D_{-bj_3,-bj_2,-bj_1}^{-b\rho_3,-b\rho_2,-b\rho_1}
$$
where $C$ stands for the boundary three point function in the
$SL(2,\mathbb{R})$ WZNW model, $D$ is the boundary three point
function in Liouville field theory; a boundary operator is
labelled by its spin $j$, and its left and right boundary
conditions $\rho_1$ and $\rho_2$. The labels here are integers
submitted to the conditions \footnote{These conditions make the
 four point function of boundary operators regular at $z=x$.}
$$
\rho_2=\rho_1+j_1-n,  \ \ j_3=j_2+j_1-m, \ \ n,m \in \mathbb{N}
$$
We do not infer that these two models are equivalent:
%(in the pseudosphere case, the metric is euclidean, whereas the
%space $AdS_3$ is equipped with a lorentzian metric):
it was shown in \cite{ZAZA} in the pseudosphere case, the possible
boundary conditions are parametrized by two positive integers
$(s,t)$, and if $s>1$, the one point function does not have any
usual classical limit. In the WZNW model considered here, the
boundary conditions are parametrized by one positive integer only,
and the one point function does have a smooth classical limit.
The situation is very analogous to Liouville theory on the pseudosphere
with $s=1$.

\section*{Acknowledgments}
B.P. would like to thank S.~Theisen and Al.~B.~Zamolodchikov for discussions, and in
particular N.~Drukker for his interest in this work. Work supported by EU under
contract HPRN-CT-2000-00122.


\begin{thebibliography}{99}

\bibitem{Sta}
S.~Stanciu,
``D-branes in an AdS(3) background'',
JHEP {\bf 9909} (1999) 028, hep-th/9901122.


\bibitem{BaPe}
C.~Bachas and M.~Petropoulos,
``Anti-de-Sitter D-branes'',
JHEP {\bf 0102} (2001) 025, hep-th/0012234.


\bibitem{PeRi}
M.~Petropoulos and S.~Ribault,
 ``Some remarks on anti de Sitter D-branes'', JHEP {\bf 0107} (2001) 036,
 hep-th/0105252

\bibitem{Ri}
S.~Ribault, ``$AdS_2$ D-branes in $AdS_3$ spacetime'', hep-th/0207094

\bibitem{Ba}
C.~Bachas
``Asymptotic symmetries of $AdS_2$ branes'', hep-th/0205115


\bibitem{RaRo}
A.~Rajaraman and M.~Rozali,
``Boundary states for D-branes in $AdS_3$'', hep-th/0108001.


\bibitem{HiSu}
Y.~Hikida and Y.~Sugawara,
``Boundary states of D-branes in $AdS_3$ based on discrete series'', hep-th/0107189.


\bibitem{PS}
A.~Parnachev, D.~A.~Sahakyan, ``Some remarks on D-branes in $AdS_3$'',
JHEP \textbf{0110} (2001) 022, hep-th/0109150


\bibitem{LOPT}
P.~Lee, H.~Ooguri, J.~W.~Park and J.~Tannenhauser,
``Open strings on $AdS_2$ branes'',
Nucl.\ Phys. {\bf B610} (2001) 3, hep-th/0106129.



\bibitem{GKS}
A.~Giveon, D.~Kutasov and A.~Schwimmer,
``Comments on D-branes in $AdS_3$'',
Nucl.\ Phys. {\bf B615} (2001) 133, hep-th/0106005.

\bibitem{LOP}
P.~Lee, H.~Ooguri and J.~W.~Park, ``Boundary states for $AdS_2$
branes in $AdS_3$'', Nucl. Phys. {\bf B632} (2002) 283-302,
hep-th/0112188.




\bibitem {PST}B.~Ponsot, V.~Schomerus, J.~Teschner,
 ``Branes in the euclidean $AdS_3$'', JHEP \textbf{0202} (2002) 016, hep-th/0112198


\bibitem{FZZ}V.~A.~Fateev, A.~B.~Zamolodchikov and Al.~B.~Zamolodchikov, ``Boundary Liouville field theory I'',
hep-th/0001012

\bibitem{ZAZA}A.~B.~Zamolodchikov, Al.~B.~Zamolodchikov, ``Liouville field theory on a pseudosphere'', hep-th/0101152

%\bibitem{Wakimoto}
%M.~Wakimoto, ``Fock representation of the algebra $A_1^{(1)}$'', Comm. Math. Phys. {\bf 104}, 605 (1986)\\
%A.~B.~Zamolodchikov (unpublished)\\
%D.~Bernard and G.~Felder, ``Fock representation and BRST cohomology in $SL(2)$ current algebra'', Comm. Math. Phys. {\bf 127}, 145 (1990)

\bibitem{P}B.~Ponsot, ``Monodromy of solutions of the Knizhnik-Zamolodchikov equation: $SL(2,\mathbb{C})/SU(2)$ WZNW model'', Nucl.~Phys.~\textbf{B642} (2002), 114-138, hep-th/0204085


%\bibitem{dot}

%M.~Bershadsky and H.~Ooguri, ``Hidden $sl(n)$ symmetry in conformal theory'',
%Comm. Math. Phys \textbf{126} 49 (1986)\\
%V.~S.~Dotsenko, ``Solving the $su(2)$  conformal field theory with the Wakimoto
%free field representation'',  Nucl. Phys. \textbf{B358} 547 (1990)


\bibitem {PT3}B.~Ponsot, J.~Teschner, ``Boundary Liouville Field Theory: Boundary three point function'',  Nucl. Phys. \textbf{B622} 309 (2002), hep-th/0110244






















\end{thebibliography}
\end{document}